\documentclass[conference]{IEEEtran}
\IEEEoverridecommandlockouts
\usepackage{amsmath, amssymb}    
\usepackage{bm}                 
\usepackage{graphicx}           
\usepackage{enumitem}           
\usepackage{tcolorbox}          
\usepackage{url} 
\usepackage{amsmath,amsfonts}
\usepackage{cite}
\usepackage{amsmath,amssymb,amsfonts}
\usepackage{algorithmic}
\usepackage{graphicx}
\usepackage{textcomp}
\usepackage{xcolor}
\usepackage{amsthm}
\usepackage{graphicx}     
\usepackage{float}
\usepackage{siunitx}
\DeclareSIUnit{\dBW}{dBW}
\usepackage{graphicx}
\usepackage{caption}
\usepackage{subcaption}
\usepackage{float} 

\usepackage{algorithm}
\usepackage{tcolorbox}   
\usepackage{booktabs}
\usepackage{booktabs} 
\usepackage{array}    

\usepackage[letterpaper, top=0.75in, bottom=1.04in, left=0.63in, right=0.63in]{geometry}
\def\BibTeX{{\rm B\kern-.05em{\sc i\kern-.025em b}\kern-.08em
    T\kern-.1667em\lower.7ex\hbox{E}\kern-.125emX}}

\begin{document}

\title{Modeling and Mitigation of 7.125-7.40 GHz Terrestrial Network RFI on the Passive Earth Exploration Satellite Service in 6.725-7.125 GHz Band \\
{\footnotesize }
\thanks{This work was supported in part by the National Science Foundation (NSF) under Grants SWIFT-2332637 and SWIFT-2128077.}
\thanks{\copyright\ 2026 IEEE. Personal use of this material is permitted. Permission from IEEE must be obtained for all other uses, in any current or future media, including reprinting/republishing this material for advertising or promotional purposes, creating new collective works, for resale or redistribution to servers or lists, or reuse of any copyrighted component of this work in other works. Accepted at the 2026 IEEE International Symposium on Spectrum Innovation (DySPAN).}
}


\author{
\IEEEauthorblockN{
Md Toufiqur Rahman\textsuperscript{*}\thanks{\textsuperscript{*}These authors contributed equally to this work.}, 
Hariharan Venkat\textsuperscript{*}, 
Chung-Tse Michael Wu, 
Ivan Seskar, 
Narayan B. Mandayam
}
\IEEEauthorblockA{
Wireless Information Network Lab (WINLAB), Rutgers University, New Jersey, USA \\
\texttt{\{mtrahman,hvenkat,ctmwu,seskar,narayan\}@winlab.rutgers.edu}
}
}

\maketitle

\begin{abstract}
The 7.125--7.4~GHz band is attractive for next generation  Terrestrial Network (TN) deployments owing to the large bandwidths available and favorable propagation characteristics. Furthermore, recent U.S. Presidential actions have cleared the usage of this band for 6G by stipulating relocation of federal incumbents that share this band. However, this deployment can only be successful if we can also guarantee coexistence of these networks with existing incumbents operating in adjacent bands. This paper presents a comprehensive analysis of the Radio Frequency Interference (RFI) caused by the proposed TNs in the 7.125--7.4~GHz band at passive Earth Exploration Satellite Service (EESS) sensors that operate in the adjacent 6.725--7.125~GHz band. Using TN base stations (BSs) equipped with filtennas (filtering antennas) as well as transmit precoders for RFI mitigation, we carry out an RFI analysis that accounts for increasing BS deployments in the contiguous U.S. over a 10 year period from 2030 to 2040. We also characterize the size of the guard bands needed to protect the EESS sensors for different BS deployment densities. With appropriate filtenna and precoder design, our results reveal that a 100 Mbps increase in the rate requirements of the TN users results in an RFI increase of roughly 2.45~dB at the EESS sensors. For a 25~MHz Guard Band, simulations show that in 2030, there is no significant RFI for user rates upto 500~Mbps. 
However, the same systems in 2040 would cause RFI that is around 4~dB above the ITU mandated threshold for passive EESS sensors. 
This would need to be countered by $(a)$ increasing Guard Band width to $35$~MHz, or 
$(b)$ by reducing the user data rate requirements to $300$~Mbps.



\end{abstract}


\section{Introduction}
In the search for new candidate spectrum in next-generation Terrestrial Networks (TNs), the 7.125--7.4~GHz band stands out. As part of the newly emerging Frequency Range 3 (FR3) spectrum (7.125--24.25~GHz), this band is argued to combine the best of both worlds from traditional sub-6 GHz bands and mmWave bands. Operations in this band can unlock the higher capacities needed for next generation use cases since it offers significantly more contiguous bandwidth than existing sub-6~GHz allocations~\cite{ITU_M2160, 5GAmericas2024}. It's propagation characteristics are $(i)$ favorable in comparison to propagation in the mmWave band, and $(ii)$ comparable to the 6~GHz band which allows for the reuse of existing macro-cell architectures~\cite{Prop_Study_7GHz, Ericsson_7GHz_2024}. This band has also been tapped for its potential in Integrated Sensing and Communications (ISAC) applications. The wavelengths allow for massive Multiple Input Multiple Output (MIMO) arrays that are small enough for mobile User Equipment (UEs), while also being large enough to provide the required spatial degrees of freedom for sensing applications~\cite{ISAC_Study_7GHz, Huawei_ISAC_2024}. Due to these advantages, the United States has identified parts of this band for study through congressional action~\cite{One_Big_Beautiful_Bill_2025,FCC_TAC_Charter3_Propagation_2024}. The recent Presidential Memorandum titled ``Winning the 6G race” discusses the relocation of federal incumbents in the 7.125 -- 7.4~GHz range to make way for 6G development, thus reaffirming the United States’ interest in using this band for terrestrial 6G~\cite{Trump_6G_Race_2025}.

Interest in using 7.125--7.4~GHz frequencies for International Mobile Telecommunication (IMT) is not new. Global momentum for this band was established at the World Radiocommunication Conference (WRC) 2023 where the International Telecommunications Union (ITU) adopted resolutions 813 and 256. These documents designated the identification of the 7.125--8.4 GHz band (or parts thereof) for next generation IMT as one of the main agendas of WRC-27~\cite{ITU_WRC23_Res813,ITU_WRC23_Res256}. In the US, the National Telecommunications and Information Administration's (NTIA) National Spectrum Strategy (NSS) document from November 2023 also motivated the study of this portion of spectrum for wireless broadband~\cite{NTIA_NSS_2023}. 

However, the ITU resolutions and the NTIA NSS also stressed the 
need for studies regarding the coexistence of the proposed IMT 
deployments with incumbent services in and adjacent to its band 
of operation. Among these incumbents, the passive Earth Exploration 
Satellite Service~(EESS) represents a uniquely 
vulnerable service — its protected allocations span 
6.725--7.125~GHz, placing it in direct spectral adjacency to 
the proposed 7.125--7.4~GHz IMT band~\cite{fcc_2023_preliminary}. 

Passive EESS satellites carry highly sensitive radiometers to detect minute variations in Earth’s microwave brightness temperature~\cite{ITUR_RS1861, NASA_AMSR2_Technical}. A primary example is JAXA’s GCOM-W1 mission, which utilizes the Advanced Microwave Scanning Radiometer (AMSR) sensors. These sensors deliver geophysical products across two retrieval levels: at Level-1, calibrated brightness temperatures~(L1B/L1R) serve as the foundational observable; at Level-2, these are translated into Integrated Water Vapor~(TPW), Cloud Liquid Water~(CLW), Precipitation~(PRC), Sea Surface Temperature~(SST), Sea Surface Wind Speed~(SSW), Sea Ice Concentration~(SIC), Snow Depth~(SND), and Soil Moisture Content~(SMC) — products that collectively underpin global numerical weather prediction, ocean state estimation, and long-term climate monitoring~\cite{ITUR_RS2017}. Critically, the passive nature of EESS radiometry imposes a fundamental constraint: unlike active services, passive sensors cannot filter or distinguish natural Earth emissions from anthropogenic interference, making any in-band or adjacent-band emission source a direct threat to measurement integrity.

The threat to the passive EESS band is not hypothetical. This band is currently under co-channel pressure from mass-deployed Wi-Fi~6E/7 infrastructure. The 6~GHz band~(5.925--7.125~GHz) comprises four Unlicensed National Information Infrastructure (U-NII) segments: U-NII-5, U-NII-6, U-NII-7, and U-NII-8. This entire range is now allocated for unlicensed Radio Local Area Network (RLAN) use. Within this allocation, Low Power Indoor (LPI) access points are authorized to operate at a Power Spectral Density (PSD) of $+$5~dBm/MHz without the requirement for Automated Frequency Coordination (AFC)~\cite{dogan2024spectrum}. Consequently, co-channel emitters in U-NII-7 and U-NII-8 now directly overlap the EESS allocation at 6.725--7.125~GHz. Current literature has quantified the aggregate interference from this RLAN environment to geostationary satellite incumbents. These studies report a worst-case aggregate Interference-to-Noise (I/N) ratio of $-$25.1~dB under conservative deployment assumptions~\cite{yoza2020spectrum}. However, these analyses focus primarily on Fixed-Satellite Service (FSS) uplink receivers, representing a notable research gap regarding the impact on passive EESS sensors. Furthermore, the primary attenuation factor for indoor LPI emissions toward spaceborne sensors—Building Entry Loss (BEL)—is  statistically distributed parameter per ITU-R Rec.~P.2109. Since median BEL values span 24.5--36.8~dB depending on building type and elevation angles~\cite{yoza2020spectrum}, a non-trivial upward propagation path remains for RLAN emitters to interfere with passive EESS sensors. Whether aggregate RLAN emissions already constitute a limiting 
interference floor for passive EESS sensors remains an open 
question beyond the scope of this work. Notably, the passive EESS
frequency reference itself is under revision, with WRC-27 Agenda 
Item~1.19 tasked with evaluating potential new allocations in 
the 4\,200--4\,400\,MHz and 8\,400--8\,500\,MHz bands~\cite{wrc23_res674}.


This work addresses the structurally distinct and 
more severe threat posed by the proposed IMT deployments at 
7.125--7.4~GHz. Unlike Wi-Fi~6E/7 LPI devices which are 
low-power, predominantly indoor, and subject to BEL, terrestrial IMT networks deploy high-power base stations (BSs) outdoors with fixed infrastructure, directional antennas, and continuous high-duty 
cycle operation. The spectral proximity of the EESS 
channel to the proposed IMT band makes 
these sensors acutely susceptible to Out-Of-Band~(OOB) 
emissions and receiver overload. Even low-level OOB 
interference corrupts Level-1 brightness temperature 
retrievals, with cascading degradation across all dependent 
Level-2 geophysical products — directly compromising the 
weather forecasting and climate monitoring services that EESS enables. This underscores the urgency of the rigorous coexistence analysis presented in this paper, as mandated by WRC-23 Resolutions~256 and~813~\cite{ITU_WRC23_Res813,ITU_WRC23_Res256} for IMT coexistence, and Resolution~674~\cite{wrc23_res674} 
for EESS allocations.


\section{Related Works}
The evolution of modern TNs toward FR3 has catalyzed a plethora of works on spectrum coexistence and sharing. Broadly, these works address coexistence through either mutually exclusive resource access or simultaneous operation facilitated by interference-mitigation architectures. Initial frameworks for terrestrial-satellite coexistence often relied on deterministic spatial or temporal separation, adhering to the mutually exclusive resource access paradigm. For instance, practical IMT and EESS sharing in the 7--8 GHz band has been studied using Real-Time GeoFenced Spectrum Sharing~\cite{Eichen2024}. While such methods ensure deterministic protection, they inherently introduce downtime in the terrestrial network. In the case of~\cite{Eichen2024}, the authors noted that while IMT services maintain operations approximately 99.9\% of the time, the remaining 0.1\% "paused" time necessitates the invocation of Open RAN (O-RAN) techniques to migrate subscribers to alternative frequencies. 

To refine these rigid boundaries, recent research has also explored Dynamic Exclusion Zones (DEZ) enabled by Machine Learning (ML). The authors in \cite{MandavaGlobecom2025} assessed DEZ and power control for coexistence between 5G networks and Digital Video Broadcasting-Satellite (DVB-S2) stations. Utilizing SDR-based emulation on the COSMOS testbed, the authors demonstrated that while DEZ strategies can restore DVB-S2 link performance to within 82\%--99\% of ideal conditions, they still impose constraints on terrestrial network capacity and continuity. The inherent "on-off" nature of these exclusion-based solutions requires careful analysis of network downtime and subscriber handovers, motivating the search for architectures that permit simultaneous spectral access.

On the other hand, a more efficient approach would involve simultaneous access and coexistence through the design of RFI-suppresion algorithms and dedicated hardware systems.  Conventional RFI suppression typically relies on the concatenation of standalone band-pass filters with antenna arrays. However, this modular approach introduces high insertion loss, degrades the system noise figure, and impedes the miniaturization required for dense 6G deployments. To overcome these limitations, the \emph{filtenna}--an integration of filter and antenna structures has emerged as a critical architecture \cite{FiltennaDesign}. 

Resource allocation algorithms specifically tailored for filtenna-equipped nodes have shown significant gains in reducing spectral leakage into passive bands \cite{FiltennaResource}. To leverage this hardware selectivity, recent work has integrated interference-aware constraints into downlink spatial filtering. Building on foundational beamforming techniques for power minimization \cite{Bjornson2014}, our group has shown RFI-aware precoders preserve the tractability of perfect-CSI QoS beamforming while introducing a coexistence-driven power budget~\cite{2025FNWF}. 

Due to the speculative nature of these deployments and the RFI they may cause, many works have focussed on obtaining better models for the IMT network structures in the future and quantifying the impact that RFI could pose for the satellite receivers. For instance, stochastic geometry has been used to model the aggregate effect of terrestrial emitters on satellite receivers~\cite{Koosha2022}. Meanwhile other works have developed comprehensive interference models for the coexistence of 6G systems and passive sensing systems~\cite{Testolina2023}. Authors in~\cite{NWP_Impact} evaluate the degradation of Numerical Weather Prediction (NWP) accuracy as a function of 5G mmWave deployment density \cite{NWP_Impact}.

Our previous work~\cite{2025FNWF} established an optimal precoder formulation for filtenna-equipped systems but was limited to single-link interference modeling. This work attempts to use the proposed optimal precoders and built a more accurate and realistic model of RFI at EESS sensors due to the adoption of TNs operating at the proposed 7.125-7.4 GHz band over the 10 year span from 2030-2040. This paper,

\begin{itemize}
    \item Accounts for aggregate RFI caused by all the BSs in a satellite's footprint. BS deployment densities over a 10 year period are obtained based on modeling realistic technology adoption rates using a Gompertz model for all metropolitan counties in the contiguous United States~\cite{golparvar2024dyspan}. 
    \item Provides accurate estimates of the RFI caused by a base station  for different data rate demands of the UEs it services. This is done because performance metrics like UE data rates in the 7.125--7.4~GHz next generation systems are currently undecided.
    \item Invokes optimal RFI-aware spatial multi-filtering for DownLink (DL) from our group’s previous work~\cite{2025FNWF}. The precoders from that work are used here to ensure that the terrestrial BSs meet user rate requirements with least transmit power while ensuring RFI threshold limits are met.
    \item Uses accurate models for the passive EESS constellation which consists of 5 different sensors ($\mathrm{B1}$ through $\mathrm{B5}$), each with different orbital parameters obtained from the official ITU-R RS.1861-1 document~\cite{ITUR_RS1861}. The maximum permissible RFI threshold value for passive EESS sensors of -166~dBW/ 200~MHz is also obtained from~\cite{ITUR_RS1861}. 
    \item Explores the requirements and specifications of a ``guard band” in the proposed TN band. The EESS band (6.725-7.125 GHz) and the proposed 7.125-7.4 GHz band have no guard bands in between, which could cause RFI values to exceed the ITU mandated thresholds. This paper studies RFI at the EESS for multiple guard band intervals and provides the necessary recommendations to protect the EESS based on the RFI thresholds from~\cite{ITUR_RS1861}.
\end{itemize}

\section{System Model}
\label{SysMod}
\begin{figure}[h!]
    \centering    \includegraphics[width=.80\columnwidth]{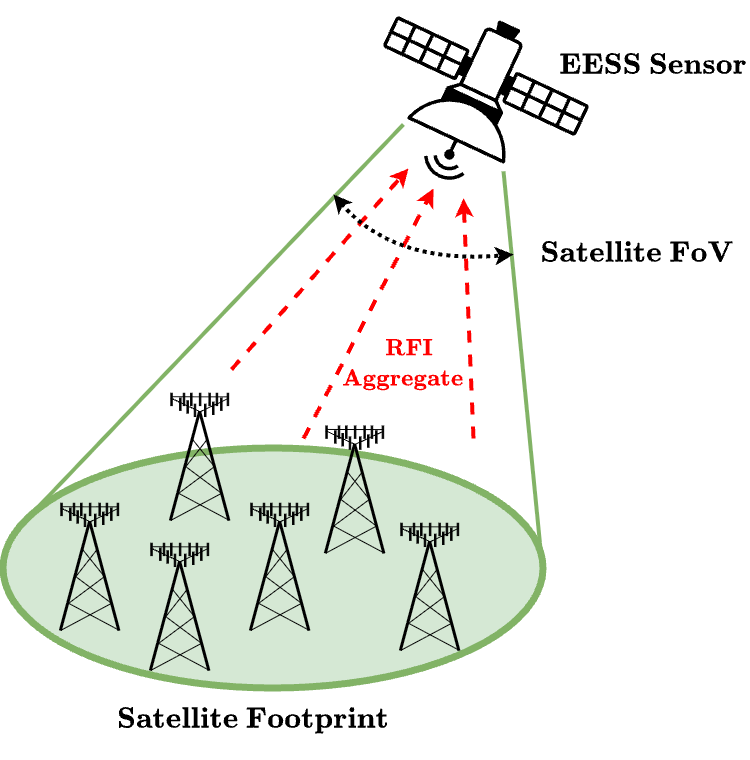}
    \caption{Aggregate RFI is computed over all Base Stations in the satellite footprint}
    \label{MultipleBS}
\end{figure}

The adoption of the proposed 7.125--7.4~GHz TN is studied in this work. Specifically, BS deployments are modeled over the ten year interval from 2030-2040 in all metropolitan counties of the contiguous United States. As shown in Fig~\ref{MultipleBS}, the aggregate RFI contribution of all BSs within the satellite footprint is considered in order to gauge the effect of the IMT deployments in this band.

RFI aggregation consists of computing the RFI due to a single BS in a cell and multiplying that by the number of base stations in the satellite footprint. This assuming that all base stations are equidistant to EESS sensor. Analyzing RFI due to a single BS is carried out in Section~\ref{SingleBSSec} using the results from our group’s previous work~\cite{2025FNWF} which introduced optimal RFI-aware precoding algorithms for a filtenna-equipped BS. Meanwhile, the number of BSs deployed is modeled in Section~\ref{GompertzSection} with the help of a Gompertz model as seen in~\cite{golparvar2024dyspan}.

\subsection{Single Cell Model}\label{SingleBSSec}
\begin{figure}[h!]
    \centering    \includegraphics[width=.96\columnwidth, trim=3.5cm 16cm 1cm 2.5cm, clip]{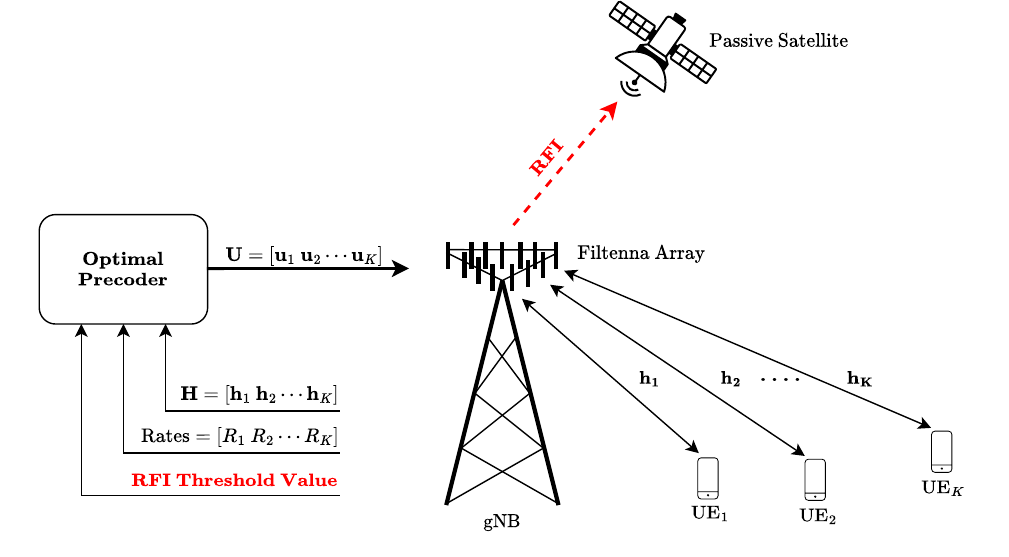}
    \caption{System Model for a single cell}
    \label{SystemModelSingleBS}
    \vspace{-1em}
\end{figure}

We consider a downlink (DL) multi-ser Multiple Input Single Output (MU-MISO) system where a single BS equipped with $N$ transmit antennas simultaneously serves $K$ single-antenna users, as illustrated in Fig.~\ref{SystemModelSingleBS}. Users may be located at different distances from the BS, and each user $k$ has a minimum rate requirement $R_{\min,k}$.

The BS is assumed to have \emph{perfect} instantaneous channel state information (CSI) of all users, i.e., it knows the channel matrix
$\mathbf{H}=[\mathbf{h}_1,\ldots,\mathbf{h}_K]\in\mathbb{C}^{N\times K}$ exactly, where $\mathbf{h}_k\in\mathbb{C}^{N\times 1}$ denotes the small-scale fading channel from the BS to user $k$. Let $\mathbf{x}\in\mathbb{C}^{N\times 1}$ denote the transmitted signal vector. Using linear precoding, the BS transmits
\begin{equation}
\mathbf{x}=\sum_{k=1}^{K}\mathbf{w}_k\,s_k,
\label{eq:x_signal_perfect}
\end{equation}
where $\mathbf{w}_k\in\mathbb{C}^{N\times 1}$ is the precoding vector for user $k$ (including beam direction and power allocation), and $s_k$ is the information symbol for user $k$ satisfying $\mathbb{E}[|s_k|^2]=1$ and $\mathbb{E}[s_k]=0$. The average BS transmit power is
\begin{equation}
P_{\mathrm{tx}}=\mathbb{E}\!\left[\|\mathbf{x}\|_2^2\right]
=\sum_{k=1}^{K}\big\|\mathbf{w}_k\big\|_2^2.
\label{eq:ptx_perfect}
\end{equation}

The received baseband signal at user $k$ is given by
\begin{equation}
y_k=\sqrt{g_k}\,\mathbf{h}_k^{H}\mathbf{w}_k\,s_k
+\sqrt{g_k}\sum_{j\neq k}\mathbf{h}_k^{H}\mathbf{w}_j\,s_j
+n_k,
\label{eq:yk_perfect}
\end{equation}
where $g_k\in\mathbb{R}_{+}$ denotes the large-scale channel gain (path loss and shadowing), and $(\cdot)^H$ denotes the Hermitian transpose. The additive noise is modeled as $n_k\sim\mathcal{CN}(0,\sigma_n^2)$, where the noise power is
$\sigma_n^2 = k_B T B$ with $k_B$ the Boltzmann constant, $T$ the receiver noise temperature (K), and $B$ the system bandwidth (Hz). Accordingly, the instantaneous SINR at user $k$ is
\begin{equation}
\mathrm{SINR}_k=
\frac{g_k\big|\mathbf{h}_k^H\mathbf{w}_k\big|^2}
{\sigma_n^2+g_k\sum_{j\neq k}\big|\mathbf{h}_k^H\mathbf{w}_j\big|^2}.
\label{eq:sinr_perfect}
\end{equation}

\begin{figure}[h!]
    \centering
    \includegraphics[width=\columnwidth]{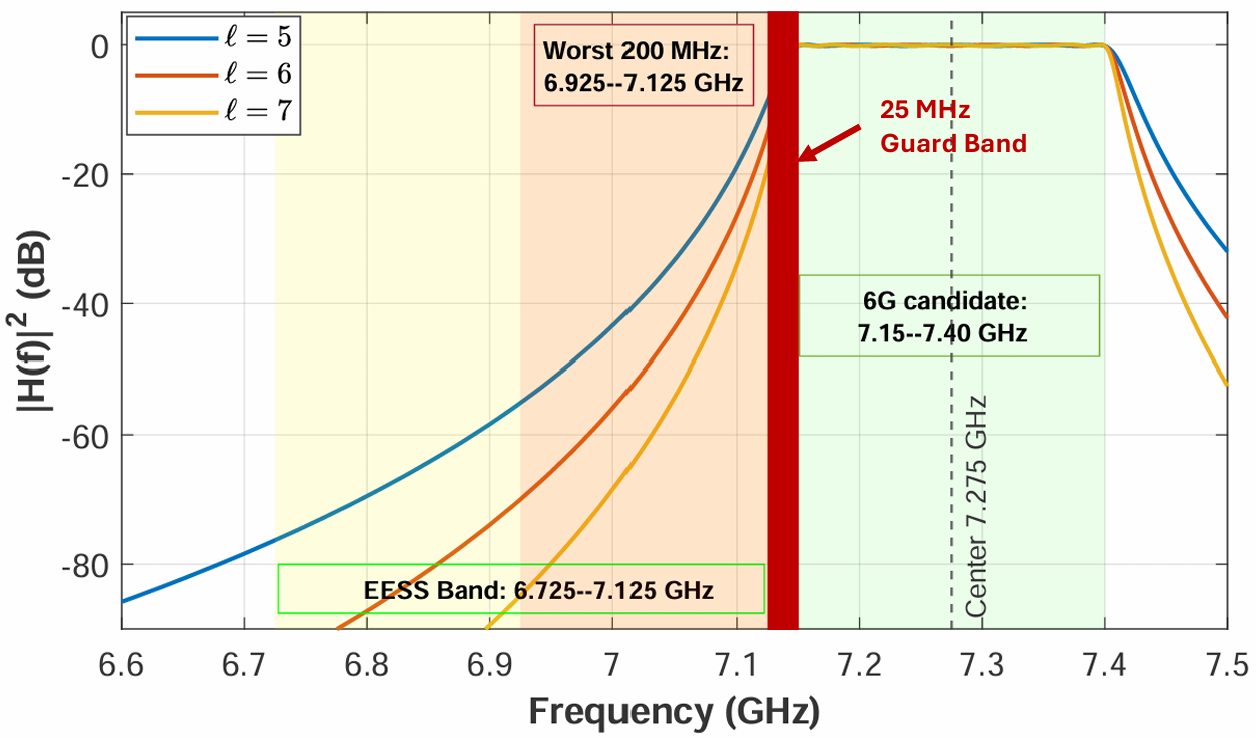}
    \caption{Chebyshev bandpass filter responses for different filter orders $\ell$ along with important bands highlighted}
    \vspace{-1em}
    \label{fig:filter_response}
\end{figure}

In order to jointly capture transmit-power consumption and its RFI implications, we decompose the BS consumed power into four components as seen in \cite{majumdar2022resource,2025FNWF}. The first component is the radiated transmit power, denoted by $P_{\mathrm{Tx}}$. The second component accounts for data-converter (DAC/ADC) power, modeled as a bandwidth-dependent fraction of the transmit power, $\alpha(B)P_{\mathrm{Tx}}$, where $\alpha(B)=\alpha_0+\alpha_1B$ increases with the signal bandwidth $B$ \cite{majumdar2022resource}. Third, we include the additional hardware power drawn by filtennas through a factor $\beta(l)P_{\mathrm{Tx}}$, where $\beta(l)=\beta_0+\beta_1l$ captures the increased circuit complexity as the number of coupled-resonator stages (filter order) $l$ grows \cite{vosoughitabar2024filtenna1}. Increasing the filter order $\ell$ sharpens the spectral roll-off and improves out-of-band leakage suppression for Chebyshev-prototype bandpass designs. At the same time,  higher-order implementations typically incur greater hardware and power overhead \cite{majumdar2022resource,2025FNWF}. Fig.~\ref{fig:filter_response} illustrates the magnitude-squared responses $|H(f)|^2$ for several filter orders $\ell$, and highlights the passive EESS sensor band (6.725--7.125~GHz), the 200~MHz window (6.925--7.125~GHz) in the EESS band with maximum RFI (due to its proximity), and the 6G candidate band (7.125--7.400~GHz) with an explicit 25 MHz guard band shown. Owing to this guard band, the IMT band spans from 7.15--7.4 GHz. All discussions henceforth assume a guard band of 25~MHz (unless mentioned otherwise), filtenna order $\ell = 7$, and a terrestrial band from 7.15--7.4~GHz resulting in a bandwidth of 250~MHz.

To quantify the resulting RFI contribution, we model the fraction of transmit power that leaks into adjacent bands as $\delta(\ell)P_{\mathrm{Tx}}$, where $\delta(\ell)$ is a decreasing function of $\ell$ due to improved spectral containment at higher filter orders \cite{majumdar2022resource}. Collectively, the total power $P_{\mathrm{Tot}}=P_{\mathrm{Tx}}\big(1+\alpha(B)+\beta(l)+\delta(l)\big)$. A maximum tolerable satellite interference level $I_{\text{sat, max}}$ is also imposed. After propagation over the BS--SAT interference path with gain (including path loss) $g_{\text{sat}}$, the resulting received RFI at the satellite is $g_{\text{sat}}\delta(l)P_{\text{Tx}}$. Therefore, for fixed $l$ (and fixed $\delta(l)$), controlling average RFI is equivalent to controlling the BS transmit power. In particular, minimizing $P_{\text{Tx}}$ via the choice of precoders $\{\mathbf{w}_k\}$ simultaneously reduces the total consumed power $P_{\text{Tot}}$ and the induced satellite RFI. If no precoder can satisfy both the user QoS requirements and the satellite RFI limit, the design is declared infeasible for that $l$. 

If $\gamma_k$ denotes the target SINR for user $k$, then for a fixed varactor stage $\ell$ the RFI-aware power-minimization problem is formulated as
\vspace{-0.4em}
\begin{equation}
\label{eq:rfi_power}
\begin{aligned}
\min_{\{\mathbf{w}_k\}_{k=1}^{K}} \quad & \sum_{k=1}^{K}\|\mathbf{w}_k\|^{2} \\
\text{s.t.}\quad 
& \mathrm{SINR}_k \ge \gamma_k, \quad \forall k,\\
& g_{\mathrm{sat}}\,\delta(\ell)\sum_{k=1}^{K}\|\mathbf{w}_k\|^{2} \le I_{\mathrm{sat,max}} .
\end{aligned}
\end{equation}

The RFI constraint can be mapped to an equivalent transmit-power budget
$P_{\mathrm{sat,max}} = I_{\mathrm{sat,max}}/(g_{\mathrm{sat}}\delta(\ell))$,
yielding an active per-run power budget $P_{\mathrm{sum,max}}=\min\{P_{\mathrm{BS}},P_{\mathrm{sat,max}}\}$.
When $P_{\mathrm{sat,max}}<P_{\mathrm{BS}}$ (RFI-limited regime), tightening $I_{\mathrm{sat,max}}$
reduces $P_{\mathrm{sum,max}}$ and increases infeasibility/outage for a given $\gamma$.
When $P_{\mathrm{sat,max}}\ge P_{\mathrm{BS}}$ (BS-limited regime), the RFI constraint is inactive
and further increasing $I_{\mathrm{sat,max}}$ provides no additional gain.Therefore, the problem in \eqref{eq:rfi_power} can be written equivalently as
\vspace{-0.4em}
\begin{equation}
\label{eq:rfi_power_min_with_active_budget}
\begin{aligned}
\min_{\{\mathbf{w}_k\}_{k=1}^{K}} \quad & \sum_{k=1}^{K}\|\mathbf{w}_k\|^{2} \\
\text{s.t.}\quad 
& \mathrm{SINR}_k \ge \gamma_k, \quad \forall k,\\
& \sum_{k=1}^{K}\|\mathbf{w}_k\|^{2} \le P_{\mathrm{sum,max}}.
\end{aligned}
\end{equation}

The problem in \eqref{eq:rfi_power_min_with_active_budget} is originally non-convex due to the fractional quadratic nature of the SINR constraints. Mathematical methods to reformulate this constraint into a convex second-order cone (SOC) constraint, and subsequent solution methods are discussed in~\cite{Bjornson2014, bengtsson2001optimal, 2025FNWF}.

As shown in~\cite{majumdar2022resource,2025FNWF}, for the minimum-power formulation
in~\eqref{eq:rfi_power_min_with_active_budget}, the optimal beamformers
$\{\mathbf{w}_k^\ast\}_{k=1}^K$ satisfy the SINR constraints with equality,
$\mathrm{SINR}_k=\gamma_k$, resulting in the minimum transmit power required for the downlink. 

\subsection{Base Station Deployment Model} 
\label{GompertzSection}
As discussed above, RFI aggregation involves accounting for all BSs in the satellite footprint area. To obtain this, we model the number of BSs deployed in every metropolitan county in the contiguous United States. If the footprint area of the satellite is given by $A_{Sat}$, the area of a county by $A_{County}$, and the number of base stations in the county by $N_{BS}$, then the number of base stations from a county in the footprint of a satellite,

\begin{equation}
	N_{\text{BSs in Footprint from County}} = \left \lfloor\left(\dfrac{A_{Sat} \cap A_{County}}{A_{County}} \right) N_{BS}\right\rfloor.
\end{equation}

The implicit assumption made here is that the base stations are uniformly distributed over the county area. We model the base station deployments in the 7.15--7.4~GHz band from a tentative starting date for deployment (assumed to be 2030) and over a 10 year span. To model the adoption trajectory, we first model the percentage of a county’s population who adopt the technology, $N_{u}$. Then, we compute the number of base stations as $N_{BS} = (N_{u} * \text{Max Data Demand per User}) / {(\eta * B)}$, where $\eta$ represents the spectral efficiency and $B$ is the bandwidth of operation. To model $N_{u}$, we use a Gompertz diffusion model and estimate its parameters from historical U.S.\ landline internet subscription penetration (subscriptions per 100 people, 1998--2023) using nonlinear least squares (NLS) \cite{MeadeIslam1998,golparvar2024dyspan}.


\begin{figure}[h!]
    \centering
    \includegraphics[width=.90\columnwidth]{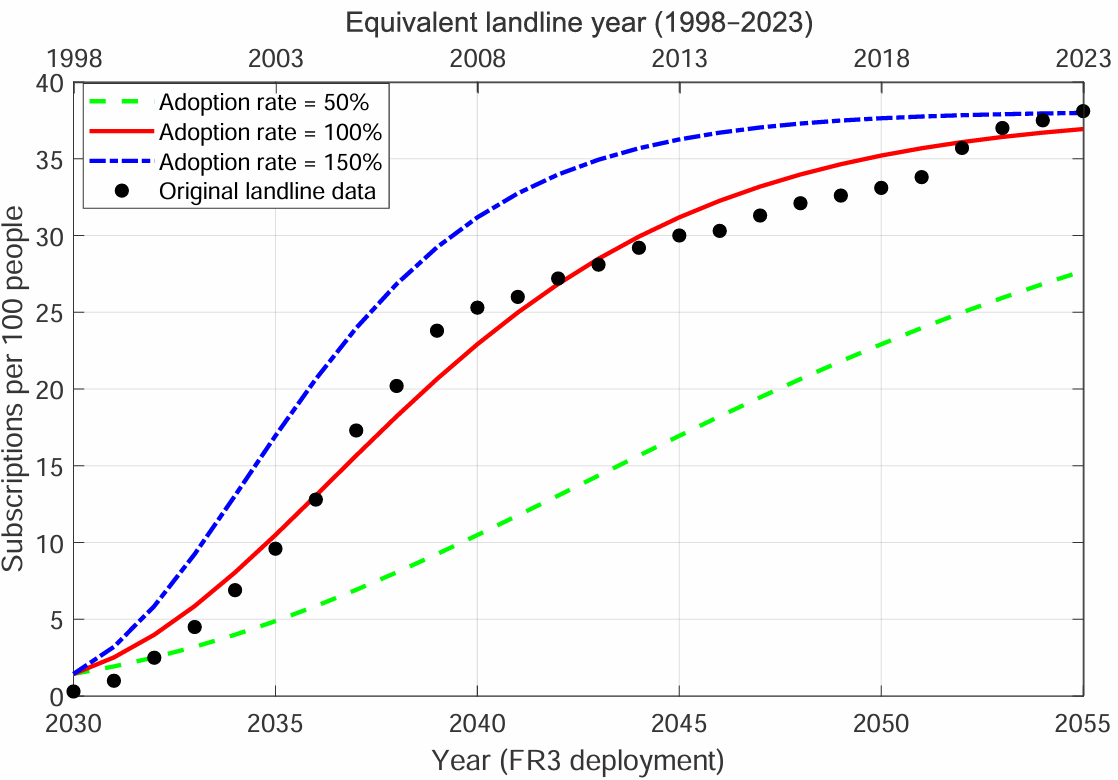}
    \caption{FR3 deployment trajectory from a Gompertz model fitted to historical landline data for three adoption rate scenarios.}
    \label{fig:fr3_adoption}
\end{figure}

\noindent The model is given by

\begin{equation}
    Y(t) = b_1 \exp\!\bigl(-b_2 \exp(-b_3 (t))\bigr),
    \label{eq:gompertz}
\end{equation}
where $Y_t$ denotes subscription penetration (subscriptions per 100 people) at year index $t$, and $b_1$, $b_2$, and $b_3$ are learnable parameters. Here, $b_1$ represents the asymptotic saturation level, while $b_2$ and $b_3$ govern the displacement and growth rate of the adoption curve, respectively. In our fit, we obtain
\(
(b_1,\, b_2,\, b_3) \;=\; (38.100,\, 3.272,\, 0.186)
\). The fitted curve serves as a shape template for the projected FR3 deployment 
trajectory, illustrated in Fig.~\ref{fig:fr3_adoption} for three adoption 
rate scenarios (50\%, 100\%, and 150\%), with a sensitivity analysis provided 
in the Appendix.
In the numerical results that follow, we assume a maximum spectral efficiency ($\eta$) per base station of 50 bps/Hz with 100\% adoption rate. The list of metropolitan counties and their populations is obtained from the most recent rural-urban continuum codes published by the United States Department of Agriculture in 2023~\cite{USDARUCC}. County populations are assumed to remain unchanged over the observation period.

\subsection{EESS Model}
\label{EESSMod}
Based on the ITU-R RS.1861/RS.1861-1 documents~\cite{ITUR_RS1861}, we evaluate the terrestrial BS$\rightarrow$EESS interference link for five representative EESS sensors $\{\mathrm{B1,B3,B4,B5,B7}\}$ using the altitude and incidence-angle parameters. These sensors were chosen due to the proximity of their frequency of operation to the TN band of interest. For each sensor $k$, the BS--sensor slant range $D_k$ is obtained from the RS.1861-1 spherical-Earth geometry, and the corresponding free-space path loss is computed at $f=6.925$~GHz as $L_{\mathrm{FSPL},k}\,[\mathrm{dB}]=92.45+20\log_{10}(f_{\mathrm{GHz}})+20\log_{10}(D_{k,\mathrm{km}})$. Beyond FSPL, we include clear-sky additional attenuation terms that are kept explicit in the link budget: a polarization mismatch loss $L_{\mathrm{pol}}=3$~dB (BS $\pm45^\circ$ dual-slant vs.\ Horizontal / Vertical sensor polarization), gaseous absorption $L_{\mathrm{atm}}=0.3$~dB following \cite{ITUR_P676}, and an Earth--space terminal clutter loss $L_{\mathrm{clut}}=5.5$~dB consistent with a representative mid-elevation pass($25^\circ$) and median location percentile ($p=50\%$) \cite{ITUR_P2108}. Denoting $L_{\mathrm{add}}=L_{\mathrm{pol}}+L_{\mathrm{atm}}+L_{\mathrm{clut}}$, the total propagation loss is $L_{\mathrm{tot},k}=L_{\mathrm{FSPL},k}+L_{\mathrm{add}}$. Additional impairments such as rain and cloud attenuation can be incorporated when required using \cite{ITUR_P618}. 

The transmit-side gain on the BS$\rightarrow$EESS path is defined as the BS radiation gain toward the sensor look direction. 
Since the BS beams are primarily steered toward terrestrial UEs while the satellite sensors are typically observed off-axis, the coupling is generally sidelobe-dominated. Since 6G systems are estimated to employ massive MIMO arrays with sharp beams and low sidelobe levels (SLLs) along with advanced capabilities like null-steering~\cite{Nokia_6G_2025}, we assume a conservative antenna gain of -10 dB directed towards the satellite. This is in accordance with the antenna discrimination requirements established in Recommendation ITU-R M.2101-0 \cite{ITUR_M2101} that accounts for the SLLs of a massive MIMO array. 

The resulting net link gain (partial link budget) for sensor $k$ is then $\mathcal{G}_k \triangleq G_{\mathrm{Tx}} + G_{\mathrm{Rx},k} - L_{\mathrm{tot},k}$ [dB], where $G_{\mathrm{Rx},k}$ is the sensor maximum receive gain from RS.1861/RS.1861-1. Table~\ref{tab:eess_sensor_netgain} reports the computed geometry and losses at $f=6.925$~GHz, with slant ranges spanning $1066$--$1610$~km and total losses spanning $178.6$--$182.2$~dB. Among the five sensors, $\mathrm{B5}$ yields the largest (least negative) net gain, $\mathcal{G}_{\mathrm{B5}}=-128.79$~dB, due to its substantially higher receive gain of 51.5~dBi and distinct proximity to the proposed TN band. Therefore, $\mathrm{B5}$ is selected as the worst-case victim sensor for the subsequent RFI-limited performance evaluation. To ensure spectral compatibility, the received interference power at sensor $k$, $P_{\mathrm{RFI},k}$,
must not exceed the applicable EESS(passive) protection threshold; in particular,
$P_{\mathrm{RFI},k}\le -166$~dBW when integrated over a 200~MHz reference bandwidth, consistent with
the performance and interference criteria ~\cite{ITUR_RS2017}.

\begin{table}[t]
\centering
\caption{BS$\rightarrow$EESS net coupling for adjacent-band coexistence with 5G BS operation at 7.25--7.4\,GHz.}

\label{tab:eess_sensor_netgain}
\resizebox{\columnwidth}{!}{%
\begin{tabular}{c c c c c c c c}
\hline
\textbf{Sensor} & \textbf{Altitude $H$} & \textbf{Incidence $i$} & \textbf{Slant $D$} & \textbf{$G_{\mathrm{Rx}}$} & \textbf{Net Link Gain} & \textbf{Channel span} & \textbf{Footprint Area}\\
\textbf{(ID)}   & \textbf{(km)}         & \textbf{(deg)}         & \textbf{(km)}      & \textbf{(dBi)}            & \textbf{(dB)}          & \textbf{(GHz)}   & \textbf{(sq. km)}     \\
\hline
B1 & 705.00 & 55.00 & 1124.2 & 38.8 & $-145.28$ & 6.750--7.100 & 3225 \\
B3 & 830.00 & 65.00 & 1610.3 & 35.5 & $-151.70$ & 6.750--7.100 & 11690\\
B4 & 699.60 & 55.00 & 1116.2 & 40.6 & $-143.41$ & 6.750--7.100 & 2170 \\
B5 & 820.00 & 55.00 & 1292.9 & 51.5 & $-133.79$ & 6.725--7.125 & 209\\
B7 & 665.96 & 55.00 & 1066.2 & 40.6 & $-143.02$ & 6.750--7.100 & 1881 \\
\hline
\end{tabular}%
}
\end{table}

\subsection{Leakage Model}
To quantify adjacent-band emissions into the passive sensors, we define a leakage factor $\delta_s(\ell)$ as the \emph{fraction} of BS transmit power that falls into a worst-case 200\,MHz victim window within sensor $s$'s receive allocation. For a Chebyshev Type-I bandpass design of order (stage) $\ell$ with frequency response $H_\ell(f)$, the victim window $W_s$ is selected as the 200\,MHz sub-band closest to the BS operating band, since this region is typically most affected by transition-band leakage. The leakage factor is then computed as the BS-bandwidth-normalized integral of the filter power response over $W_s$, i.e.,
$\delta_s(\ell)=\frac{1}{B_{\mathrm{BS}}}\int_{f\in W_s}\!\left|H_\ell(f)\right|^2\,df$,
where $B_{\mathrm{BS}}$ denotes the occupied BS bandwidth. In practice, we evaluate this integral numerically using trapezoidal integration over a dense frequency grid, so that $\delta_s(\ell)$ can be directly interpreted as a power-leakage fraction.

As discussed in Section~\ref{EESSMod}, sensor $\mathrm{B5}$ is expected to have the worst (maximum) RFI impact due to TN deployments. As a result, we will show all plots for this sensor and attempt to compute optimal system parameters which limit the RFI / 200~MHz at this sensor to less than -166~dBW as per ITU recommendations~\cite{ITUR_RS2017}. This will then automatically imply that all other sensors will have RFI within range. 

The OOB leakage analysis we use conforms to the operating band unwanted emissions (OBUE) mask and spurious
emission limits defined in 3GPP\,TS\,38.104 \cite{3gpp_ts38104} for 5G\,NR base stations operating at frequencies above
\SI{1}{GHz}. For a maximum transmit power 
of \SI{-5}{dBW} and the adopted guard band of \SI{25}{MHz}, the power spectral density of the leaked power through a 7-th order Chebyshev bandpass filter of 250~MHz bandwidth is found to be -18.3\,dBm/MHz at the edge of the EESS band (7.124-7.125~GHz). This satisfies the Category\,A spurious limit of $-13$\,dBm/MHz by $5.3$\,dB/MHz, confirming that the proposed 6G--satellite coexistence
framework meets the 3GPP\,TS\,38.104 emission requirements.


\subsection{Terrestrial Network Channel Model}
We consider a downlink MU-MISO system with $N=256$ transmit antennas at the BS.  For each Monte Carlo realization, user locations are generated by sampling the BS--user 2D distances uniformly within an annulus bounded by an exclusion radius $r_{\min}=10$~m and a cell radius $R_{\text{cell}}=150$~m. Each link is then assigned a Line of Sight (LOS) or Non Line of Sight (NLOS) state using the UMi--Street Canyon LOS probability model specified by the 3GPP channel standard~\cite{3gpp38901}. Subsequently, large-scale path loss is computed according to the UMi--Street Canyon outdoor model, including the breakpoint-based LOS formulation and the NLOS rule defined as the maximum of the LOS loss and the additional NLOS loss term; log-normal shadow fading is applied with scenario-dependent standard deviations (LOS and NLOS) as specified in the same standard \cite{3gpp38901}. To obtain a simplified narrowband channel suitable for precoder evaluation under perfect CSI, small-scale fading is modeled as spatially i.i.d.\ Rayleigh fading across the BS antennas, and each user channel vector is scaled by the corresponding large-scale gain (path loss and shadow fading). A transmit gain of 15~dB is assumed from the base station to the terrestrial users. 

\section{Coexistence Case Study}
Recall the following parameters used in the simulations below. The TN band houses a 25~MHz guard band (unless otherwise specified), yielding 250~MHz bandwidth for TN operations. TN BSs are equipped with 256 filtennas ($\ell = 7$) which provide a conservative antenna gain of 15~dB towards their users and a -10~dB gain towards the satellite. A transmit power threshold of -5~dBW is set. The population of counties is assumed to be constant from 2030 to 2040. All RFI plots refer to the EESS sensor $\mathrm{B5}$.

\subsection{Modeling Base Station Deployment across the Contiguous United States}
\begin{figure}[h!]
    \centering    \includegraphics[width=\columnwidth, trim=1cm 4cm 19.5cm 1cm, clip]{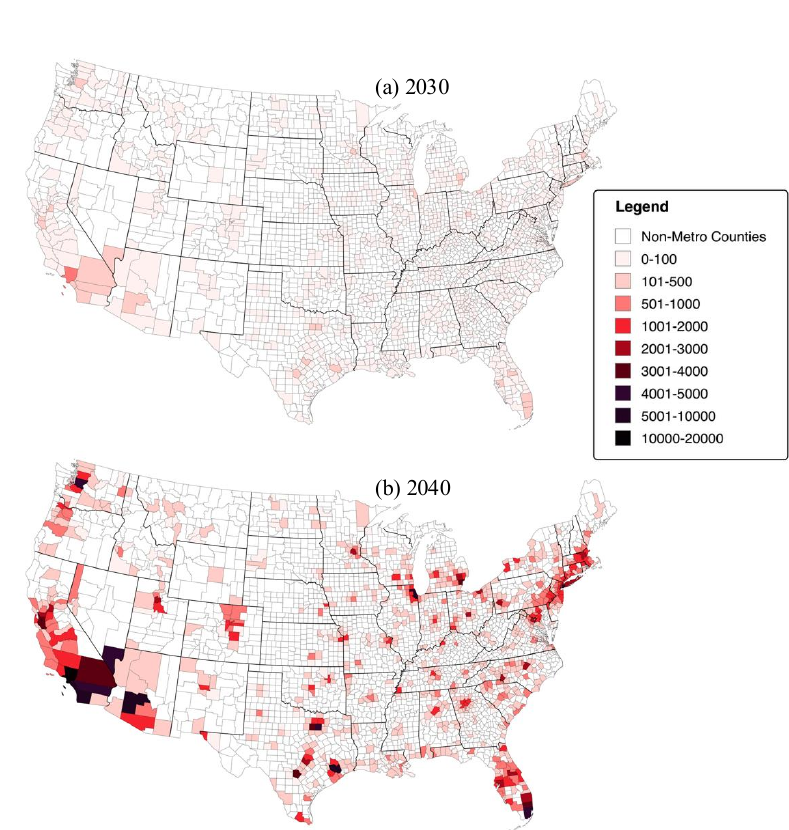}
    \caption{Predicted Base Station Number in each metropolitan county of the contiguous United States in (a) 2030 and (b) 2040}
    \label{fig:MultipleBSModelUS}
\end{figure}

As discussed above, the Gompertz model is used to compute the number of base stations per metropolitan county in the United States. The results (assuming a user rate requirement of 100~Mbps) of the simulation are shown in Fig.~\ref{fig:MultipleBSModelUS}.

\begin{figure}[h!]
    \centering
    \includegraphics[width=0.8\columnwidth]{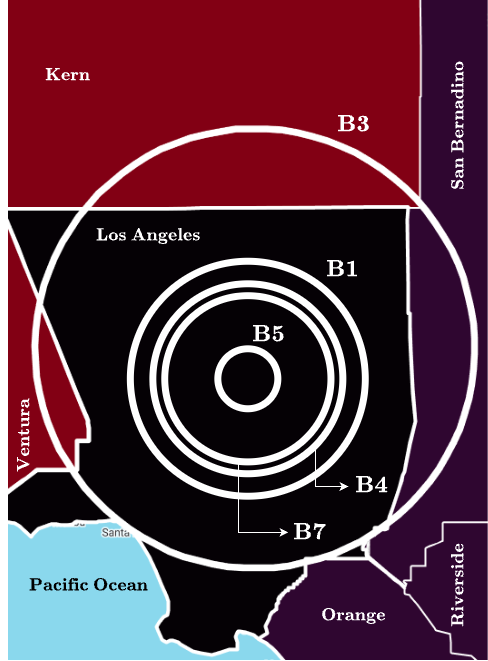}
    \caption{A zoom in on LA county (black) with ``worst case" satellite footprints (with largest base station numbers) shown. Adjoining counties are shown.}
    \label{fig:LAZoom}
\end{figure}

Upon closer examination, it is clear that the county with maximum number of base stations that fall within a satellite footprint is Los Angeles (LA) county, owing to its high population of roughly 10 Million. This work assumes uniform density of base stations within a county. The area of counties also need to be accounted for in deciding a ``worst" case county. Since most satellite footprints are smaller than the LA county area (as seen in Fig.~\ref{fig:LAZoom}), the effect of the BS distribution is pronounced (as opposed to say, New York county which has high BS density but small area and lower density surrounding counties). Thus, all RFI values reported are for this ``worst case" footprint of EESS sensors over LA County.

\begin{table*}[h!]
\centering
\caption{Aggregate RFI at the satellite (dBW/200\,MHz). Each entry reports $\{2030, 2035, 2040\}$.}
\label{table:rfi_full}
\footnotesize
\setlength{\tabcolsep}{3pt}
\begin{tabular*}{\textwidth}{@{\extracolsep{\fill}} l ccccc}
\toprule
\textbf{Sensor} & \textbf{100 Mbps} & \textbf{200 Mbps} & \textbf{300 Mbps} & \textbf{400 Mbps} & \textbf{500 Mbps} \\
\midrule
\textbf{B1} & $-198.5, -188.5, -185.0$ & $-194.9, -184.9, -181.4$ & $-192.5, -182.5, -179.0$ & $-190.7, -180.7, -177.2$ & $-188.9, -178.9, -175.4$ \\
\textbf{B3} & $-199.3, -189.3, -185.8$ & $-195.7, -185.7, -182.2$ & $-193.4, -183.4, -179.8$ & $-191.5, -181.5, -178.0$ & $-189.8, -179.8, -176.3$ \\
\textbf{B4} & $-198.3, -188.3, -184.8$ & $-194.7, -184.7, -181.2$ & $-192.4, -182.4, -178.9$ & $-190.6, -180.6, -177.0$ & $-188.8, -178.8, -175.3$ \\
\textbf{B5} & $-185.6, -175.6, -172.1$ & $-181.8, -171.8, -168.3$ & $-179.5, -169.5, -166.0$ & $-177.6, -167.6, -164.1$ & $-176.0, -166.0, -162.5$ \\
\textbf{B7} & $-198.6, -188.6, -185.1$ & $-195.0, -185.0, -181.5$ & $-192.6, -182.6, -179.1$ & $-190.8, -180.8, -177.3$ & $-189.0, -179.0, -175.5$ \\
\bottomrule
\end{tabular*}
\end{table*}

\subsection{Impact of Guard Band on RFI }

\begin{figure}[h!]
    \centering
    \includegraphics[width=\columnwidth]{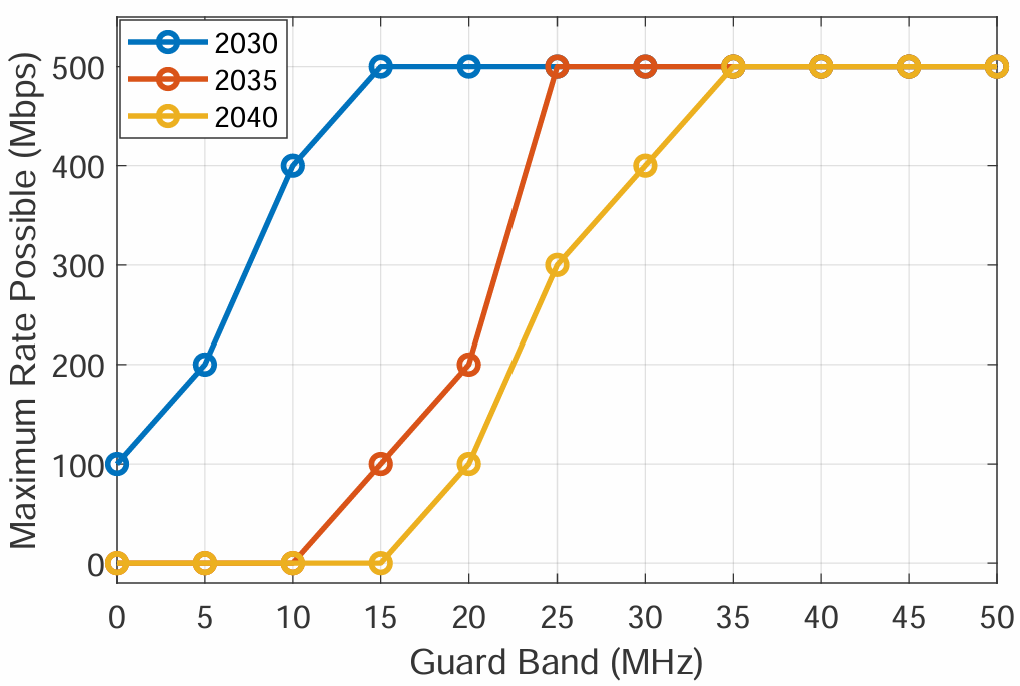}
    \caption{Maximum Rate Feasible Under RFI Constraint Across Guard Bands (2030/2035/2040)}
    \label{fig:rfi_vs_guard}
\end{figure}

\vspace{-1mm}
Coexistence analysis begins with simulating the effects of a guard band in the TN. We study TN guard band widths from 0 to 50~MHz in 5~MHz intervals and the associated RFI on the EESS sensors. As the width of the guard band increases, the bandwidth available for the TNs decreases. This implies that a larger number of BSs would be needed to maintain the same rate requirements for the users, along with a higher power requirement. At the same time, the RFI due to a single BS would be lowered due to the wider guard band. The outcome of these opposing effects is studied in Fig~\ref{fig:rfi_vs_guard}. The y-axis represents the maximum possible rate achievable (rounded to the nearest 100 Mbps, and capped at 500 Mbps) such that the aggregate RFI at the satellite does not exceed the ITU threshold. It is noted that the effect of RFI is not pronounced in 2030 and 2035 (the initial year/s) as 500 Mbps user rate requirements are met. It is to be noted that in 2035, the RFI value is -166.02 dBW/200~MHz which brings it very close to the threshold. The further densification of base station deployments in 2040 results in a loss of around 200~Mbps in the TN's maximum achievable rate if they adhere to the RFI thresholds mandated by the ITU. 

\subsection{RFI Comparison Under Perfect CSI}
\begin{figure}[t]
    \centering
    \includegraphics[width=\linewidth]{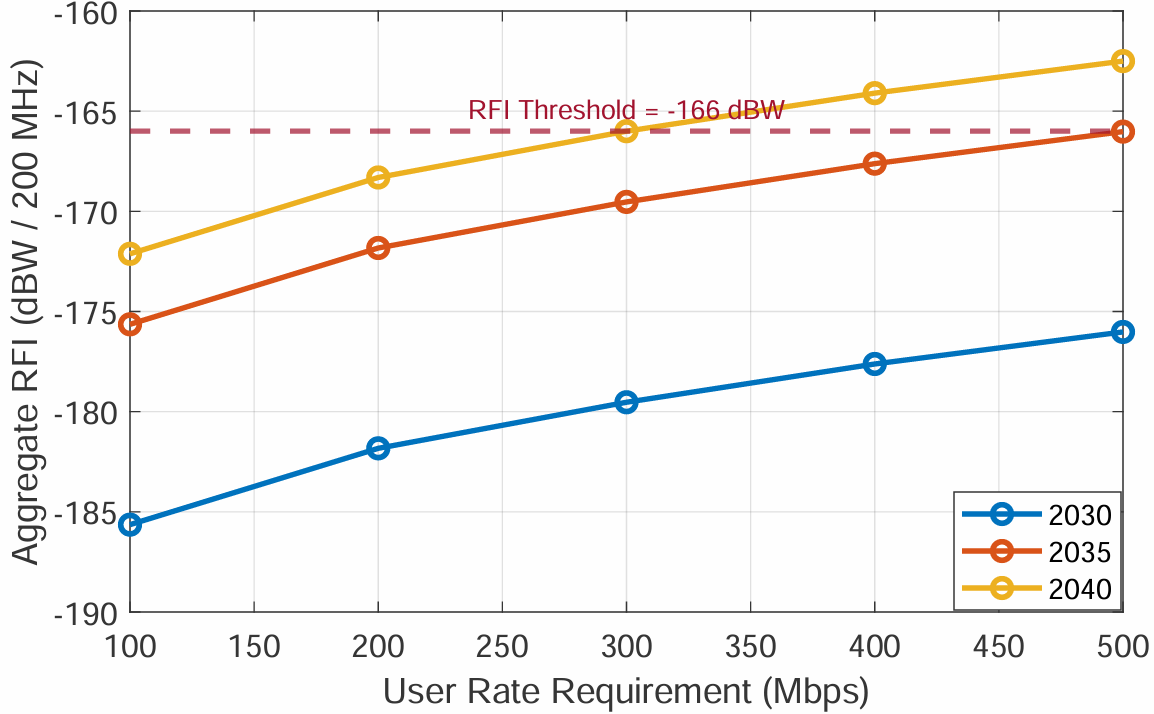}
    \caption{Aggregate RFI at the EESS ($\mathrm{B5}$) versus target downlink rate for \(\ell=7\), for 2030, 2035, and 2040.}
    \label{fig:aggRFI_B5_l6_years}
\end{figure}

The values RFI for different user rate requirements in the years 2030, 2035, and 2040 are shown here. Fig.~\ref{fig:aggRFI_B5_l6_years} shows the value of RFI as a function of the rate requirements in the TNs. As the target rate increases, the required transmit power rises and results in higher aggregate RFI. From the analysis, we note an average increase of 2.45~dB in RFI for every 100~Mbps increase in rate requirement. Additionally, for TNs with user requirements of 500~Mbps, RFI values are well below the ITU-mandated thresholds in 2030 and 2035. The same systems in 2040, however, exceed the -166 dBW / 200~MHz threshold by around 4~dB respectively. This would need to be countered by $(a)$ increasing the guard band width to 35~MHz as seen from Fig~\ref{fig:rfi_vs_guard}, or $(b)$ reducing the user data rates to 300~Mbps as seen from Fig~\ref{fig:aggRFI_B5_l6_years}. 

As seen in Table.~\ref{table:rfi_full}, the other four sensors all observe very similar RFI values despite them occupying orbits with differing orbital parameters. This may be due to the footprint area and path loss scaling as the square of orbital altitude. This way, sensors farther away have higher path losses, but cover a larger area which implies more base stations in their footprint. These inverse effects seem to cancel each other, yielding similar RFI results among the other four sensors. The sensors also all operate on the same frequency band which is further from the TN band when compared to $\mathrm{B5}$.

\section{Conclusion}
In this paper, the effect of RFI at EESS satellites due to the proposed 7.125--7.4~GHz band was studied. Base station deployment models were used to compute the exact aggregate RFI incident at the sensitive EESS sensors. Accurate ITU data was used to model link gains from the TNs to the EESS sensors. Guard band widths to protect EESS operations were studied under different deployment densities. Our simulations show that a guard band of 25~MHz is enough to guarantee 500 Mbps user rates in 2030, but by 2040 as deployments get denser, a guard band of atleast 40~MHz
is required for the same effect. RFI as a function of user rates was also studied and showed that a 100~Mbps increase in user rate requirements translates to a 2.45~dB increase in RFI at EESS sensors. 

\appendix
\section{3GPP\,TS\,38.104 Compliance Verification}
\label{appendix:leakage}


To assess sensitivity to adoption rate uncertainty, the Gompertz growth
parameter $b_3$, in \eqref{eq:gompertz}, is scaled by factors of $0.5\times$ and $1.5\times$
relative to the baseline fitted value ($b_3 = \hat{b}_3$), yielding a
slower-growth scenario (50\% adaptation rate) and a faster-growth scenario
(150\% adaptation rate), while the saturation level $b_1$ and displacement
parameter $b_2$ are held fixed at their fitted values. Table~\ref{tab:sensitivity}
reports the resulting subscriptions per 100 people at selected years.

\begin{table}[!h]
    \renewcommand{\arraystretch}{1.2}
    \caption{FR3 Adoption Sensitivity Analysis: Subscriptions per 100 People}
    \label{tab:sensitivity}
    \centering
    \begin{tabular}{@{}lccc@{}}
        \toprule
        Year & 50\% (slower) & 100\% (baseline) & 150\% (faster) \\
        \midrule
        2030 & 1.0  & 1.0  & 1.0  \\
        2035 & 5.0  & 10.0 & 17.0 \\
        2040 & 10.5 & 22.5 & 30.0 \\
        \bottomrule
    \end{tabular}
\end{table}
Simulations show that for the slower adoption rates, RFI values for $25$~MHz guard band and $500$~Mbps user rate requirement are below the ITU  mandated thresholds for 2030 and 2035, but exceed the threshold by 1~dB in 2040. However, the same systems in 2035 and 2040 for the faster adoption rates would cause RFI that is 3 and 5~dB above the ITU mandated thresholds respectively. 

\bibliographystyle{IEEEtran}
\bibliography{references}

\end{document}